\documentclass[letterpaper]{article}

\usepackage[T1]{fontenc}

\usepackage{geometry}
\usepackage{setspace}

\usepackage{achemso}
\usepackage{stix}
\usepackage{pifont}
\usepackage{graphicx}
\graphicspath{{Figs/}}
\usepackage[colorlinks,linkcolor=blue,citecolor=blue,urlcolor=blue]{hyperref}
\usepackage{float}

\usepackage{chemformula} 
\usepackage[version = 4]{mhchem} 

\DeclareUnicodeCharacter{2212}{-}

\renewcommand{\vec}[1]{\mathbfit{#1}}
\newcommand{\var}[1]{\mathsf{#1}}
\newcommand{\mat}[1]{\mathbf{#1}}

\newcommand{\scp}{\mathscr{p}}
\newcommand{\scq}{\mathscr{q}}
\newcommand{\scr}{\mathscr{r}}

\newcommand{\yes}{\ding{51}}
\newcommand{\no}{\ding{55}}

\def\a{\alpha}
\def\b{\beta}

\def\h{\eta}

\def\S{\Sigma}
\def\s{\sigma}

\def\w{\omega}
\usepackage{authblk}
\author[1]{Yaroslav Pavlyukh\textdagger}
\affil[1]{Independent researcher, Karlsruhe, Germany}
\author[2]{Fabien Bruneval\textdaggerdbl}
\affil[2]{Universit\'{e} Paris-Saclay, CEA, Service de recherche en Corrosion et Comportement des Mat\'{e}riaux,
 SRMP, Gif-sur-Yvette 91191, France}
\author[3]{Arno F\"{o}rster*}
\affil[3]{Department of Chemistry and Pharmaceutical Sciences, Vrije Universiteit, De Boelelaan 1105, 1081 HV Amsterdam, The Netherlands}
\title{Approaching Coupled Cluster Accuracy with Positive Semidefinite Vertex Corrected Self-Energies}
\date{%
  Email: \textdagger yaroslav.pavlyukh@gmail.com,
  \textdaggerdbl fabien.bruneval@cea.fr,
  *a.t.l.foerster@vu.nl
}

\begin{document}

\maketitle

\begin{abstract}
  Hedin's formalism of functional derivatives is the best-known method for systematically constructing correlated electronic theories, largely due to the success of its lowest-order self-energy expansion, the $GW$ approximation. Beyond $GW$, diagrammatic resummation schemes attempt to mix correlations simultaneously across all particle-particle and particle-hole channels. Because such a comprehensive treatment is computationally prohibitive for realistic molecular systems, a highly effective alternative is to fully account for electronic correlations in one specific channel, typically the particle-hole channel. This idea was recently implemented for molecular systems~{\setcitestyle{numbers}[\cite{forster_why_2024}]}, yielding a self-energy expressed in terms of excited-state energies and transition amplitudes from the solution of the Bethe-Salpeter equation, rather than the random phase approximation used in $GW$. While this approach is predictive and numerically efficient, it violates the fundamental positive-definiteness constraint of the electron spectral function in certain energy ranges. In this study, we resolve this physical flaw by deriving a positive semidefinite (PSD) extension of the theory using a rigorous framework based on the nonequilibrium Green's function formalism. The PSD constraint introduces new scattering channels and triplet intermediate states and restores the correct physical behavior. We demonstrate that it consistently improves quasiparticle energies across standard molecular benchmarks, with an accuracy comparable to coupled-cluster reference calculations.
\end{abstract}

The primary goal of quantum chemistry is to predict chemical properties directly from the laws of quantum mechanics, with as little empirical input as possible. While its traditional focus has been on ground-state structures and energetics, recent decades have brought a shift toward predictive tools for excited states and spectral properties~\cite{blase_bethesalpeter_2018, golze_gw_2019}. Molecules are a demanding testing ground for many-body perturbation theories~\cite{bruneval_benchmarking_2013, maggio_gw_2017, olsen_beyond_2019, lewis_vertex_2019, Mejuto-Zaera2021, marie_reference_2024, Ammar2024, Wen2024, Abraham2024, marie_algebraic-diagrammatic_2026}: a rich diversity of correlation regimes is encountered already in the simplest organic compounds~\cite{bartlett_coupled-cluster_2007, stuyver_diradicals_2019}. An equally important motivation for working in the language of Green’s functions is the versatility of the formalism~\cite{stefanucci_nonequilibrium_2025}. Wavefunction methods deliver highly accurate energies for discrete transitions in isolated molecules but struggle with continuous spectra and open quantum systems; Green’s functions handle both naturally, which makes them the building blocks of choice for simulating spectroscopies and quantum transport~\cite{flood_whence_2004, nitzan_electron_2003,evers_advances_2020,tuovinen_thermoelectric_2025}.

In this study, we continue our investigation of many-body approximations that treat electronic correlations in the particle-hole channel.\cite{forster_why_2024} Two methods are central here: the $GW$ approximation (GWA)~\cite{hedin_new_1965}, the premier Green's function (GF) framework for quasiparticle properties, and the Bethe-Salpeter equation (BSE)~\cite{salpeter_relativistic_1951, strinati_application_1988}, its highly successful counterpart for excited states. Conventionally, $GW$ is performed first, and the statically screened interaction $W_0=W(\omega=0)$ enters the BSE as a central ingredient~\cite{albrecht_ab_1998, benedict_optical_1998, rohlfing_electron-hole_2000, onida_electronic_2002}. As demonstrated in Ref.~\citenum{forster_why_2024} and references therein, this workflow can be rigorously reversed: the two-particle correlation function $L$ obtained from the BSE generates an improved approximation for the electron self-energy. The reversal is physically well-motivated. The random phase approximation (RPA), which dictates the screening in standard $GW$, describes classical metallic systems such as sodium clusters well,\cite{onida_ab_1995} but breaks down for localized excitons in wide-bandgap insulators~\cite{botti_time-dependent_2007}. Since the accuracy of the RPA is highly system-dependent while the BSE agrees robustly with experiment even for complex systems~\cite{blase_bethesalpeter_2018, Forster2022c}, upgrading the building block of the self-energy from $W$ to $L$ is a logical step.

This improvement comes at a cost. Compare the GWA self-energy $\Sigma_c(\var{1}, \var{2})=iG(\var{1}, \var{2})W(\var{1}, \var{2})$ and the BSE-based self-energy
 \begin{align}
  \Sigma_c(\var{1}, \var{2}) &=i v(\var{1}^+, \var{3}) G(\var{1}, \var{4})I(\var{4},\var{6},\var{2},\var{5}) L(\var{5},\var{3},\var{6},\var{3}),
  \label{eq:se}
\end{align}
where $\mathsf{j}\equiv(r_j, \sigma_j, t_j)\equiv(j, \sigma_j)$ and $I(\var{1}, \var{2}, \var{3}, \var{4})= i\delta \Sigma(\var{1}, \var{3})/\delta G(\var{4}, \var{2})$ is the particle-hole irreducible four-point kernel; Eq.~\eqref{eq:se} is derived in the Methods section. Integration over repeated arguments is implied throughout.
The GWA self-energy is manifestly invariant with respect to the exchange of external indices, whereas the self-energy Eq.~\eqref{eq:se} may violate this property. It is however invariant when $I$ and $L$ are exact. In practical calculations where the two-particle correlation function is obtained through the solution of a Bethe-Salpeter equation,
\begin{align}
  L(\var{1}, \var{2}, \var{3}, \var{4})&=L^{(0)}(\var{1}, \var{2}, \var{3}, \var{4}) + L^{(0)}(\var{1}, \var{5}, \var{3}, \var{6}) I(\var{6}, \var{7}, \var{5}, \var{8}) L(\var{8}, \var{2}, \var{7}, \var{4}),
  \label{eq:bse}
\end{align}
using an approximate kernel, this invariance is lost. Consequently, the rate function $\Gamma(\omega)$, the imaginary part of the self-energy, can no longer be cast into a Fermi golden rule form, and the electron spectral function $A(\omega)$ is not positive semidefinite~\cite{stefanucci_diagrammatic_2014}: in certain energy ranges, the density of states becomes unphysically negative.

A rigorous remedy can be formulated within the framework of nonequilibrium Green's functions~\cite{stefanucci_diagrammatic_2014, uimonen_diagrammatic_2015, bruneval_gw_2025}. The positive semidefinite (PSD) enforcement is not a mathematical regularization: it adds Feynman diagrams describing new physical scattering processes. Applied to the self-energy in Eq.~\eqref{eq:se}, it mixes contributions from different correlation channels and generates higher-order satellites of the kind otherwise accessible only through cumulant expansions~\cite{pavlyukh_vertex_2016, Loos2024, Kocklauner2025GWMolecules}. The most surprising outcome, however, concerns the main quasiparticle peaks. In previous applications, the PSD procedure cured spectral artifacts and enriched the satellite spectrum, but never convincingly improved quasiparticle properties such as the highest occupied molecular orbital (HOMO) energies.\cite{bruneval_gw_2025} In this work, we present extensive benchmarks showing that it does: the resulting self-energies reach an accuracy comparable to the equation-of-motion coupled-cluster method with single and double excitations (EOM-CCSD).

\section*{Results}
\label{sec:formalism}
The PSD construction of Refs.~\citenum{stefanucci_diagrammatic_2014, uimonen_diagrammatic_2015} rests on a simple structural observation: on the Keldysh contour, the imaginary part of a self-energy diagram can be partitioned into products of two half-diagrams, which play the role of scattering amplitudes. A self-energy built as a complete square of such amplitudes possesses rates in the Fermi golden rule form and, therefore, a nonnegative spectral function. Our program for this section is accordingly threefold: (i) to specify the ingredients entering the self-energy~\eqref{eq:se}, (ii) to identify its constituent half-diagrams (``Ingredients'' subsection), and (iii) to recombine them into complete squares, arriving at practical PSD approximations (``PSD self-energies'' subsection).
\subsection*{Ingredients}
\label{sec:ingredients}
\begin{figure}[t!]
  \center
  \includegraphics[width=0.8\textwidth]{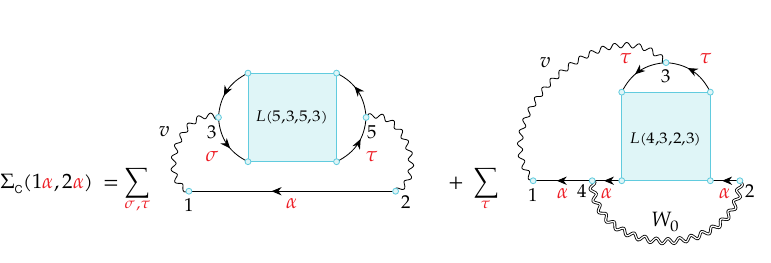}
  \begin{caption}{\small
      Spin resolved self-energy from Ref.~\citenum{forster_why_2024}. Red Greek letters indicate spins.
      \label{fig:1}}
\end{caption}
\end{figure}
We start from the universally adopted\cite{martin2016} irreducible four-point kernel
\begin{align}
  I(\var{1}, \var{2}, \var{3}, \var{4})&= \delta(\var{1},\var{3})\delta(\var{2}, \var{4}) v(\var{1}, \var{4}) - \delta(\var{1},\var{4})\delta(\var{2}, \var{3}) W_0(\var{1}, \var{3}).\label{eq:kr:L}
\end{align}
Because this kernel approximates the general four-point form by a sum of two-point interactions, the BSE, Eq.~\eqref{eq:bse} can be solved as a Casida equation. The same kernel is used in the self-energy~\eqref{eq:se}, giving rise to the direct and exchange terms $\Sigma_c=\Sigma_d+\Sigma_{2x}$. The self-energy with explicitly performed spin sums is depicted in Fig.~\ref{fig:1}. In this diagram, each vertex $\var{j}$ carries a position-spin variable $\scr_j=(r_j, \sigma_j)$ and a time variable $t_j$. In the nonequilibrium GF formalism, the times belong to the time-loop (Keldysh) contour comprising the minus ($-$) and the plus ($+$) branches. Operators on the minus branch are ordered chronologically, while operators on the plus branch are ordered antichronologically. Because interactions are time-local, $v(\var{1}, \var{2})=\delta(t_1-t_2)v(\scr_1, \scr_2)$, only the two-particle correlation function $L$ with degenerate time arguments appears in Eqs.~\eqref{eq:se} and \eqref{eq:bse}.

\begin{figure}[t!]
  \center
    \includegraphics[width=0.6\textwidth]{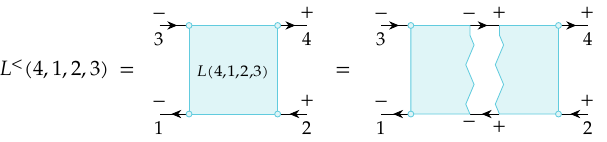}
  \hfill
  \includegraphics[width=0.37\textwidth]{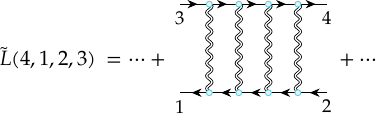}
  \begin{caption}{\small
    Partition of the two-particle correlation function with degenerate time-arguments $t_2=t_4$, $t_1=t_3$ (left). Irreducible two-particle correlation function (right). \label{fig:2}}
\end{caption}
\end{figure}

This means that $L$ adheres to the usual NEGF classification of two-time objects: $L^{T}\equiv L^{--}$,  $L^{<}\equiv L^{-+}$, $L^{>}\equiv L^{+-}$ are the time-ordered, lesser, greater components, respectively. Using the formalism of Ref.~\citenum{stefanucci_diagrammatic_2014}, a two-time lesser (or greater) object can be divided into two groups (we call them \emph{partitions}) with times on the minus and plus branches, respectively. Isolated minus or plus islands do not contribute to the total diagram. Because of the form of the kernel~\eqref{eq:kr:L}, the correlation function is \emph{one-interaction-line} reducible. The irreducible part of $L$, which we will call $\tilde{L}$,  has the form of a particle-hole ladder with instantaneous $W_0$ connecting the two fermionic lines, Fig.~\ref{fig:2} (right). Therefore, partitioning of $L$ leads to half-diagrams bridged by two internal fermionic lines as shown in Fig.~\ref{fig:2} (left), similarly to the partitioning of the polarizability ($\chi$) considered in Ref.~\citenum{uimonen_diagrammatic_2015}. Partitioning of Green's functions ($G$) was considered in Ref.~\citenum{stefanucci_diagrammatic_2014}.

\begin{figure}[t!]
  \center
  \includegraphics[width=0.95\textwidth]{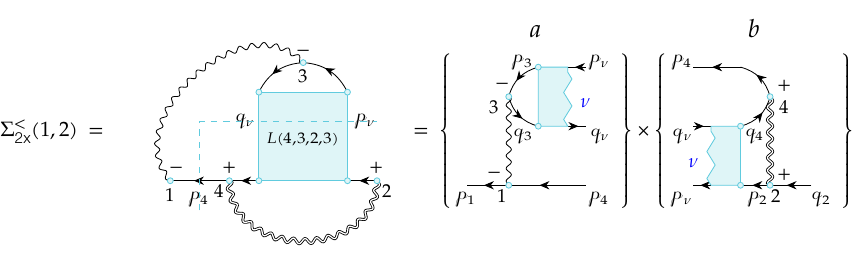}
  \begin{caption}{\small
      Partition of the self-energy diagram resulting from $I(1, 2, 3, 4)= - \delta(1,4)\delta(2, 3) W_0(1, 3)$. A dashed line denotes the border between the islands of pluses and minuses. $L$ is time-degenerate, i.e. $t_2=t_4$.
External indices are specified as $\scp_1\equiv (p,\a)$, $\scq_2\equiv (q,\a)$, where $p$ and $q$ are the orbital indices and $\alpha$ is the spin-up index. $\scp_4=(k, \a)$ ($k$ is an occupied state), and $\gamma$ label the dangling lines. For internal lines we use $\scp_2\equiv (p_2,\sigma_2),\quad \scp_3\equiv (p_3,\sigma_3),\quad\scq_3\equiv (q_3,\varsigma_3),
\quad \scp_4\equiv (p_4,\sigma_4),\quad\scq_4\equiv (q_4,\varsigma_4)$.
      \label{fig:3}}
\end{caption}
\end{figure}

With all the ingredients, we are in a position to analyze the self-energies in Fig.~\ref{fig:1}. The first (direct) term is symmetric with respect to permutation of the external lines, which is a quick way to check the PSD property. The second (exchange) diagram lacks this symmetry; moreover, its two constituent partitions (Fig.~\ref{fig:3}) are not equivalent. We treat the statically screened interaction $W_0$ as a separate interaction that appears only in certain places of the Feynman diagrams and is strictly distinct from the bare Coulomb interaction $v$. The $W_0=v$ case, leading to the so-called time-dependent Hartree-Fock (TDHF) self-energy, is considerably more complicated: the PSD procedure generates diagrams with non-degenerate $L$ (dependent on three times or three frequencies), which cannot be evaluated within the standard Casida approach. Because TDHF excited states are generally inferior to BSE ones,\cite{blase_bethesalpeter_2018,forster_why_2024} we will not pursue this scenario.

The electron self-energies in Fig.~\ref{fig:1} can be written in an explicit form highlighting their division into partitions. Let us focus on the lesser component, $\Sigma_c^<\equiv \Sigma_c^{-+}$; the greater component can be treated in a similar way. Since contour branches are specified, the lesser self-energy depends on two physical times. To simplify the notation, we will skip the explicit dependence on times and only indicate orbital/spin coordinates, i.e., $\S_c^<(\scp_1, \scq_2)$. For spin-compensated systems in a singlet ground state, the self-energy is diagonal in spin indices; therefore, it is sufficient to restrict the consideration to the spin-up case $\S_c^<(\scp_1, \scq_2)=\S_c^<(p\a, q\a)$, Fig.~\ref{fig:3}. In what follows,  we will consistently denote occupied states with indices $i,j,k$, and unoccupied states\,---\,with indices $a, b, c$. 

Let us now introduce two constituent half-diagram objects $a^\nu_{\scp_4}(p,\a)$ and $b^\nu_{\scp_4}(q,\a)$ of the lesser self-energy component. Here $\nu$ stands for the intermediate excited state and $\scp_4=(k,\s)$ is the intermediate \emph{occupied} single-particle index. We skip time arguments for brevity. With the help of standard diagrammatic rules (see Methods), they can be expressed in terms of Coulomb and screened Coulomb matrix elements and the BSE transition matrix elements $\vec{Z}^\nu$---solutions of the Casida equation:
\begin{subequations}
  \label{eq:ab:matel}
\begin{align}
  a^\nu_{k\s}(p,\alpha) & = \sum_{p_3q_3}\sum_{\sigma_3}(pk|v|q_3p_3) \vec{Z}^\nu_{p_3\sigma_3,q_3\sigma_3}\delta_{\s\a},\label{eq:a:matel}\\
  b^\nu_{k\s}(q,\alpha) & = -\sum_{p_2q_4}(p_2q|W_0|kq_4) \vec{Z}^\nu_{p_2\a,q_4\s}.\label{eq:b:matel}
\end{align}
\end{subequations}
This allows us to cast our self-energies in the sum-over-states form
\begin{align}
  \S_d^<(p\alpha,q\alpha)&
  = i \sum_{\nu\in S}\sum_k a^\nu_{k\a}(p,\alpha) a^\nu_{k\a}(q,\alpha), \label{eq:Sigma:d}\\
  \S_{2x}^<(p\alpha,q\alpha)&
  = i \sum_{\nu\in S}\sum_k a^\nu_{k\a}(p,\alpha) b^\nu_{k\a}(q,\alpha). \label{eq:Sigma:x}
\end{align}
Note that times are implicit here, and only singlet ($\nu\in S$) excited states and occupied single-particle states ($k$) contribute. The restriction to singlets is not an additional approximation: both self-energies probe $L$ only through the total (spin-summed) transition densities $\langle\nu|\delta\hat{\rho}_\a+\delta\hat{\rho}_\b|0\rangle$ appearing in its Lehmann representation, where $\delta\hat{\rho}_\s$ is the density fluctuation operator for spin $\s$. For a spin-compensated ground state, these matrix elements vanish identically for all triplet excitations (see Methods). $\S_d^<$ is already in a manifestly sum-of-squares (and therefore PSD) form, whereas $\S_{2x}^<$ is built of two different half-diagrams.

\begin{figure}[t!]
  \center
  \includegraphics[width=\textwidth]{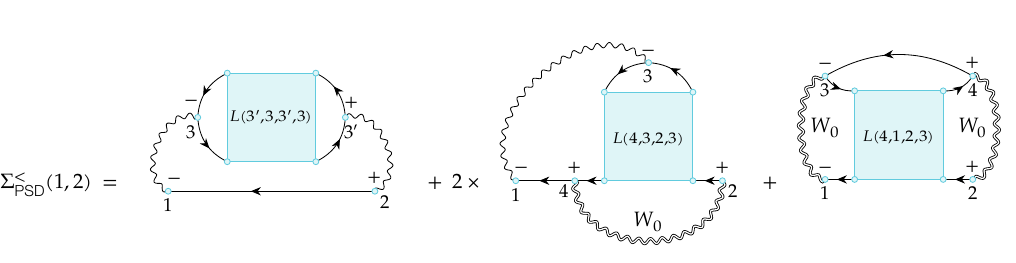}
  \begin{caption}{\small
    PSD self-energy for $W_0$ distinct from $v$ as explicitly given by Eq.~\eqref{eq:psd:gen}. \label{fig:4}}
\end{caption}
\end{figure}

The most general PSD self-energy that can be constructed from $a^\nu_{\scp_4}(p)$ and $b^\nu_{\scp_4}(q)$ reads
\begin{align}
  \S_{\text{PSD}}^{<}(p\alpha,q\alpha)&= i \sum_{\nu,k\sigma}\left[a^\nu_{k\sigma}(p,\alpha) + b^\nu_{k\sigma}(p,\alpha)\right] 
  \left[a^\nu_{k\sigma}(q,\alpha)+ b^\nu_{k\sigma}(q,\alpha)\right].\label{eq:psd:gen}
\end{align}
Here we allow for arbitrary spin multiplicity of the intermediate excited state $\nu$, and arbitrary spin $\sigma$ of the intermediate single-particle state $\scp_4$. This form is illustrated in Fig.~\ref{fig:4}. As long as $W_0$ is considered distinct from $v$, there is no double counting at any order. However, we should keep track of the double contribution of $\S_{2x}$ and a new diagram involving two $W_0$ lines and having a $T$-matrix form. Combinations of different correlation channels are typically introduced \emph{ad hoc}~\cite {Romaniello2012, Tahir2019, orlando_three_2023}, whereas here they appear naturally through the PSD construction. It is worth pointing out that the first and the third diagrams in Fig.~\ref{fig:4}, the $GW$ and $T$-matrix diagrams, respectively, are PSD by themselves, while the second is not. For the sake of classification, we define the singlet and triplet contributions to the $T$-matrix self-energy in the particle-hole channel
\begin{subequations}
  \label{se:tph}
\begin{align}
  \S_{T_\text{ph}(S)}^{<}(p\alpha,q\alpha) &= i\sum_{\nu\in S} \sum_k b^\nu_{k\a}(p,\alpha) b^\nu_{k\a}(q,\alpha), \label{eq:Sigma:Tph:S}\\
  \S_{T_\text{ph}(T)}^{<}(p\alpha,q\alpha) &= i\sum_{\nu\in T} \sum_k \left[b^\nu_{k\a}(p,\alpha) b^\nu_{k\a}(q,\alpha) \label{eq:Sigma:Tph:T}
    + b^\nu_{k\b}(p,\alpha) b^\nu_{k\b}(q,\alpha)\right].
  \end{align}
\end{subequations}
The total $T_{ph}$ self-energy consists of 
\begin{align}
  \S_{T_\text{ph}}^<(p\alpha,q\alpha)&= \S_{T_\text{ph} (S)}^{<}(p\alpha,q\alpha) + \S_{T_\text{ph} (T)}^{<}(p\alpha,q\alpha). \label{eq:Sigma:Tph}
\end{align}

\subsection*{PSD self-energies}
\label{sec:psd:se}
Starting from Eq.~\eqref{eq:psd:gen}, several practical PSD schemes can be constructed. They differ by the spin-multiplicity of the contributing excited states (only singlets $\nu \in S$ as in Eq.~\eqref{eq:Sigma:x} or singlets and triplets $\nu \in S, T$) and by their limiting behavior for $W_0\rightarrow 0$ and $W_0\rightarrow v$. In order to produce physically meaningful limits we introduce synthetic approaches, which mix several PSD self-energies---all stemming from  Eq.~\eqref{eq:psd:gen}. Using spin symmetries of the transition amplitudes $\vec{Z}^\nu$ [Eq.~\eqref{eq:Z:spin}, Methods] we arrive at simplified half-diagrams
\begin{subequations}
  \label{half:diag:expl}
\begin{align}
  a^\nu_{p_4\sigma_4}(p,\alpha) &
  = 2\sum_{p_3q_3}(pp_4|v|q_3p_3) \vec{Z}^\nu_{p_3q_3}\tfrac{1}{\sqrt{2}}\delta_{\sigma_4\a}\delta_{S_\nu,0},\label{a:diag:expl}\\
  b^\nu_{p_4\sigma_4}(q,\alpha) &
  = -\sum_{p_2q_4}(p_2q|W_0|p_4q_4) \vec{Z}^\nu_{p_2q_4}\left(\tfrac{1}{\sqrt{2}}\delta_{\sigma_4\a}\delta_{S_\nu,0} + \tfrac{1}{\sqrt{2}}\delta_{\sigma_4\a}\delta_{S_\nu,1} + \delta_{\sigma_4\b}\delta_{S_\nu,1}\right).
  \label{b:diag:expl}
\end{align}
\end{subequations}
Here, $S_\nu$ is the total spin of $\nu$. This decoupling guarantees that we only compute and store the non-vanishing spin blocks, significantly reducing the numerical complexity of the resulting self-energy routines.

\subsubsection*{Restricting sum to singlets}
\begin{figure}[t!]
  \center
  \includegraphics[width=0.6 \textwidth]{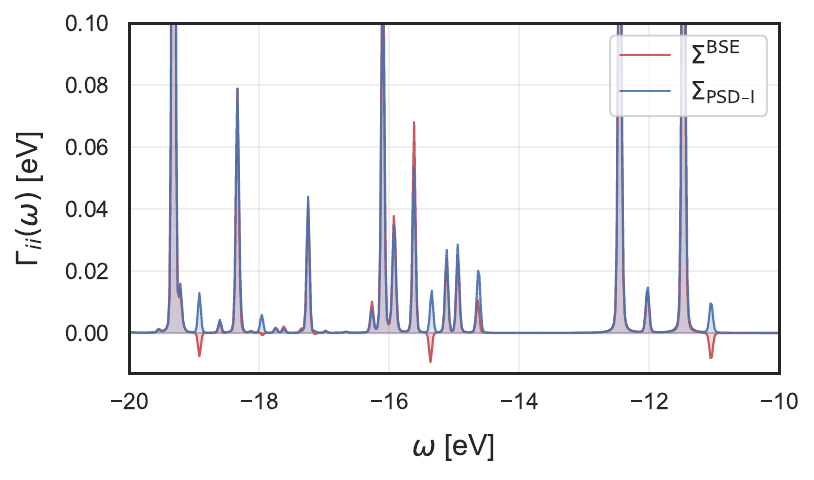}
  \begin{caption}{\small
    Rate function $\Gamma_{ii}(\w) = i[\S^>_{ii}(\w)-\S^<_{ii}(\w)]$ for $i=\text{HOMO}$ of MgO. \label{fig:5}}
\end{caption}
\end{figure}
We remind the reader that we focus here on closed-shell systems where singlet and triplet excitations are well-defined.
Let $\nu\in S$. In this case, the spin of the intermediate state is fixed ($\s=\a$) following from Eqs.~\eqref{half:diag:expl}, and we obtain our first working approximation
\begin{equation}
\begin{aligned}
  \S_{\text{PSD-I}}^{<}(p\alpha,q\alpha)&= i \sum_{\nu\in S} \sum_k\left[a^\nu_{k\a}(p,\alpha) + b^\nu_{k\a}(p,\alpha)\right] 
  \left[a^\nu_{k\a}(q,\alpha)+ b^\nu_{k\a}(q,\alpha)\right]\\
  &= \S_d^{<}(p\alpha,q\alpha) + 2\S_{2x}^{<}(p\alpha,q\alpha) +\S_{T_\text{ph}(S)}^{<}(p\alpha,q\alpha).\label{eq:PSD-I}
\end{aligned}
\end{equation}
Let us express $a^\nu_{k\a}$ and $b^\nu_{k\a}$ [given by Eqs.~\eqref{half:diag:expl}] in terms of the familiar $\vec{X}^\nu$ and $\vec{Y}^\nu$ excitation vectors. The $\vec{Z}^\nu$ matrix contains two nonzero blocks in the occupied-virtual $\vec{Z}^\nu_{ia} = \vec{X}^\nu_{ia}$ and virtual-occupied $\vec{Z}^\nu_{ai} = \vec{Y}^\nu_{ia}$ sectors. Using permutation symmetry of \emph{real} Coulomb matrix elements we obtain for $\nu\in S$
\begin{subequations}
\begin{align}
  a^\nu_{k\a}(p,\alpha) & = 2\sum_{jb}(pk|v|jb) \tfrac{1}{\sqrt{2}}\left(\vec{X}^\nu_{jb}+ \vec{Y}^\nu_{jb}\right),\\
  b^\nu_{k\a}(q,\alpha) & = -\sum_{ia}(iq|W_0|ka) \tfrac{1}{\sqrt{2}}\vec{X}^\nu_{ia} - \sum_{ia}(aq|W_0|ki) \tfrac{1}{\sqrt{2}}\vec{Y}^\nu_{ia}.
\end{align}
\end{subequations}
Inserting these expressions in Eq.~\eqref{eq:PSD-I}, and recovering the time-ordered self-energy component yields:
\begin{equation}
\begin{aligned}
  \Sigma_{\text{PSD-I},pq}^{o}(\omega) &
  = \frac{1}{2} \sum_{\nu\in S} \sum_{k} \frac{1}{\omega - \epsilon_k + \Omega_{\nu} - i\eta} \\
  &\qquad \times \sum_{ia} \left[ 2 (ai|v|qk) \left(\vec{X}^\nu_{ia} + \vec{Y}^\nu_{ia} \vphantom{\vec{Y}^\nu_{jb}} \right)
    - (ka|W_0|qi) \vec{X}^\nu_{ia} -  (ki|W_0|qa) \vec{Y}^\nu_{ia} \right] \\
  &\qquad \times \sum_{jb}\left[ 2  (bj|v|pk) \left(\vec{X}^\nu_{jb} + \vec{Y}^\nu_{jb}\right)
    - (kb|W_0|pj) \vec{X}^\nu_{jb} -  (kj|W_0|pb) \vec{Y}^\nu_{jb} \right] \;,\label{eq:Arno}
\end{aligned}
\end{equation}
where we omit explicit spin indices on $p$, $q$, and $k$.
This equation represents the simplest PSD generalisation of Eq.~(7) in Ref.~\citenum{forster_why_2024}. In Fig.~\ref{fig:5} we compare the original (BSE) and the PSD corrected rate functions for a representative molecular system (MgO). The BSE rate function develops small negative dips in a few energy windows; although their weight is minor on the scale of the dominant peaks, a negative $\Gamma(\w)$ cannot be interpreted as a scattering rate and translates into negative spectral weight. The PSD-I construction eliminates these excursions while leaving the dominant peaks essentially unchanged; the dips reappear as satellite peaks of small positive weight at nearly the same energies. This illustrates the physical content of the correction: the added $\S_{T_\text{ph}(S)}$ contribution completes the squares of the scattering amplitudes, promoting the problematic terms to genuine decay channels.

\subsubsection*{Singlets and triplets}
Let $\nu \in S\cup T$. In this case we have to add the triplet contribution: from Eqs.~\eqref{half:diag:expl} it follows $a^\nu_{k\sigma}= 0$ and $b^\nu_{k\b}= \sqrt{2} b^\nu_{k\a}$ for $\nu\in T$. We insert these into the general PSD expression~\eqref{eq:psd:gen} and obtain our second working approximation
\begin{equation}
\begin{aligned}
  \S_{\text{PSD-II}}^{<}(p\alpha,q\alpha)&= \S_{\text{PSD-I}}^{<}(p\alpha,q\alpha) + i \sum_{\nu\in T} \sum_k \left(b^\nu_{k\a}(p,\alpha)
  b^\nu_{k\a}(q,\alpha)+ b^\nu_{k\b}(p,\alpha)  b^\nu_{k,\b}(q,\alpha)\right)\\
  &= \S_{\text{PSD-I}}^{<}(p\alpha,q\alpha) +  3 i \sum_{\nu\in T} \sum_k b^\nu_{k\a}(p,\alpha)
  b^\nu_{k\a}(q,\alpha)\\
  &= \S_d^{<}(p\alpha,q\alpha) + 2\S_{2x}^{<}(p\alpha,q\alpha) +\S_{T_\text{ph}}^{<}(p\alpha,q\alpha).\label{eq:PSD-II}
\end{aligned}
\end{equation}
Explicitly we have
\begin{equation}
\begin{aligned}
  \Sigma_{\text{PSD-II},pq}^{o}(\omega) &=
\Sigma_{\text{PSD-I},pq}^{o}(\omega) \\ 
& +\frac{3}{2} \sum_{\nu\in T} \sum_{k} \frac{1}{\omega - \epsilon_k + \Omega_{\nu} - i\eta} \\
  &\qquad \times \sum_{ia} \left[ 
    (ka|W_0|qi) \vec{X}^\nu_{ia} +  (ki|W_0|qa) \vec{Y}^\nu_{ia} \vphantom{\vec{Y}^\nu_{jb}}\right]
  \cdot \sum_{jb}\left[(kb|W_0|pj) \vec{X}^\nu_{jb} +  (kj|W_0|pb) \vec{Y}^\nu_{jb} \right].
  \label{eq:PSD-II:expl}
\end{aligned}
\end{equation}
\subsubsection*{Synthetic approaches}
\label{sec:synthetic}
Let us summarise our findings. There are four classes of PSD self-energies: $\S_d$~\eqref{eq:Sigma:d}, three self-energies of the $T_{ph}$ type (\ref{eq:Sigma:Tph:S}, \ref{eq:Sigma:Tph:T} and \ref{eq:Sigma:Tph}), and two newly derived $\S_{\text{PSD-I}}$~\eqref{eq:PSD-I}, and $\S_{\text{PSD-II}}$~\eqref{eq:PSD-II}.

Setting statically screened interaction to zero in self-energies leads to $b^\nu=0$ in Eq.~\eqref{eq:psd:gen} and expressions following from this general form. In this case both, $\Sigma_{\text{PSD-I}}$ and $\Sigma_{\text{PSD-II}}$ coincide and reduce to $\Sigma_d$. If additionally we neglect $W_0$ in the kernel~\eqref{eq:kr:L} (reducing BSE to RPA), we obtain the classical $GW$ approximation.

The opposite unscreened limit $W_0\rightarrow v$ is more tricky. On one side, we cannot strictly set $W_0=v$ because our derivations rested on the fact that $W_0$ and $v$ are distinct. On the other side, if we formally perform this limit, $\S_{\text{PSD-I}}$ and $\S_{\text{PSD-II}}$ are not properly reduced to the second order perturbation theory (PT2).  This limit can be fixed by admixing $\S_d$, which is already PSD, in different proportions. Such admixture fixes the $W_0\rightarrow v$ limit at the cost of incorrect weight of the constituent $T_{ph}$ self-energies. Therefore, another strategy is to fix the bare Coulomb limit by simultaneously adding $\S_d$ and $\S_{T_{ph}}$. We emphasize that all these self-energies represent valid PSD extensions of the original $\S_{2x}$, with $\S_{\text{PSD-I}}$ being the minimal PSD form, they are compared in Tab.~\ref{tbl:cf:se}. Formal considerations alone, however, cannot single out a preferred member of this family: the limits collected in Tab.~\ref{tbl:cf:se} probe the extreme regimes of vanishing ($W_0\rightarrow 0$) and unscreened ($W_0\rightarrow v$) interaction, whereas realistic molecular systems lie in between. The relative merits of the approximations must therefore be established numerically, which is the subject of the ``Numerical Results'' subsection.
\begin{table}
  \caption{Comparison of the BSE-based PSD self-energy approximations}
  \label{tbl:cf:se}
  \centering
  \begin{tabular}{rcccc}
    \hline
    Self-energy  & $GW$ limit & PT2 limit & $T_{ph}$ limit & $\S_{2x}$ limit \\
    \hline
    $\S_{\text{PSD-I}}$ & \yes & \no & \no & \no \\
    $\frac12\left(\S_{\text{PSD-I}} + \frac34 \S_d\right)$ & \no & \yes & \no & \yes \\
    $\frac12\left(\S_{\text{PSD-I}} + \S_d\right)$ & \no & \no & \no & \yes \\
    $\frac12\left(\S_{\text{PSD-I}} + \S_d + \S_{T_{ph}(S)}\right)$ & \no & \yes & \yes & \yes\\
    \hline
    $\S_{\text{PSD-II}}$ & \yes & \no & \no & \no \\
    $\frac12\S_{\text{PSD-II}}$ & \no & \yes & \no & \yes \\
    $\frac12\left(\S_{\text{PSD-II}} + \S_d\right)$ & \no & \no & \no & \yes \\
    $\frac12\left(\S_{\text{PSD-II}} + \S_d + \S_{T_{ph}}\right)$ & \no & \yes & \yes & \yes\\
    \hline
  \end{tabular}
\end{table}
\subsection*{Numerical Results}
\label{sec:results}
\begin{figure}[t!]
  \center
  \includegraphics[trim=0cm 2.52cm 0cm 0cm, clip, width=\textwidth]{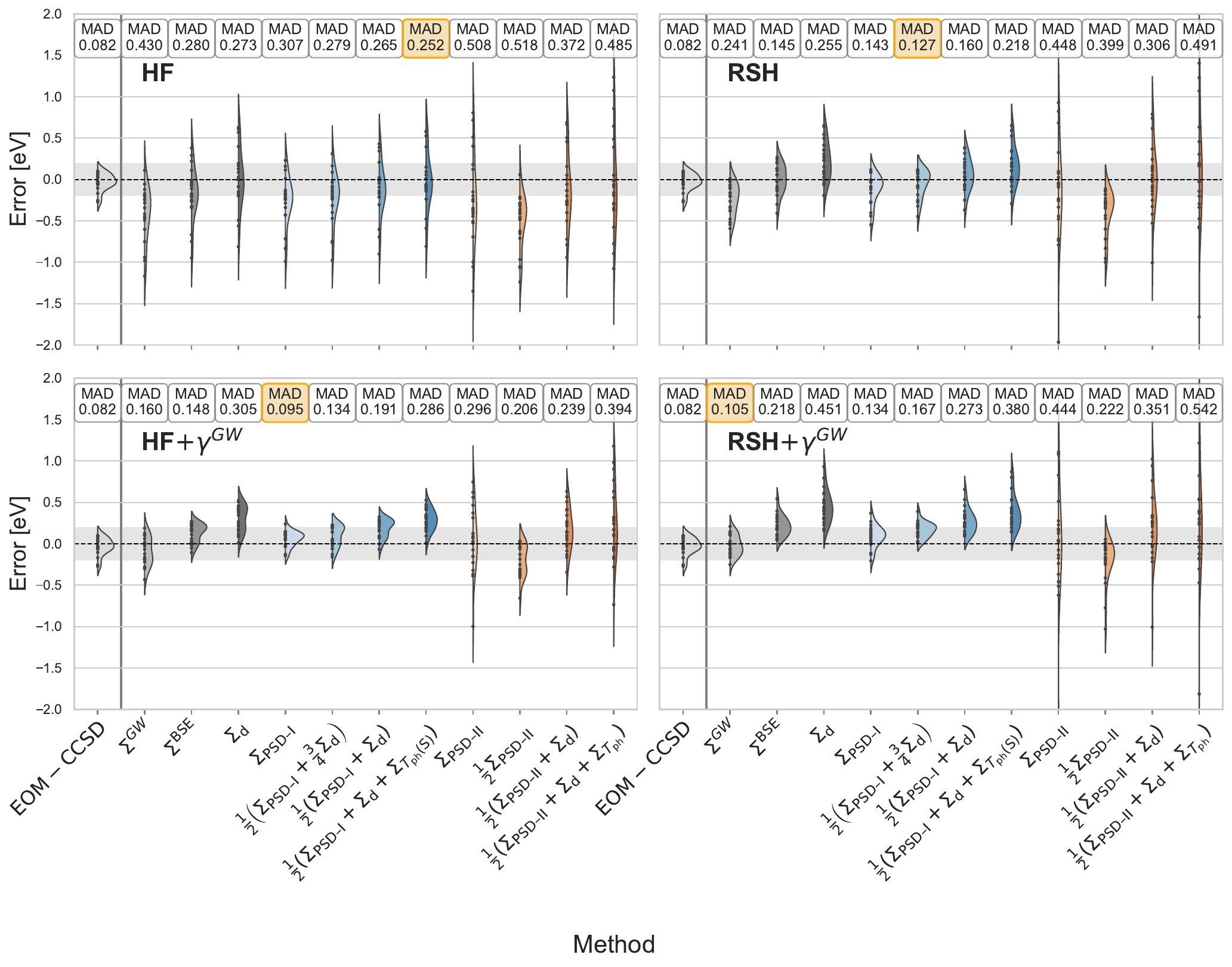}
  \begin{caption}{\small Error distribution of the HOMO energies with respect to a reference CCSD(T) $\Delta$SCF evaluation for the 23 molecules contained in the Marie and Loos set~\cite{marie_reference_2024}. Different self-energies are evaluated as one-shot perturbations on top of four different reference states.
      \label{fig:bench:homo}}
\end{caption}
\end{figure}

\begin{figure}[hbt!]
  \center
  \includegraphics[trim=0cm 2.52cm 0cm 0cm, clip, width=\textwidth]{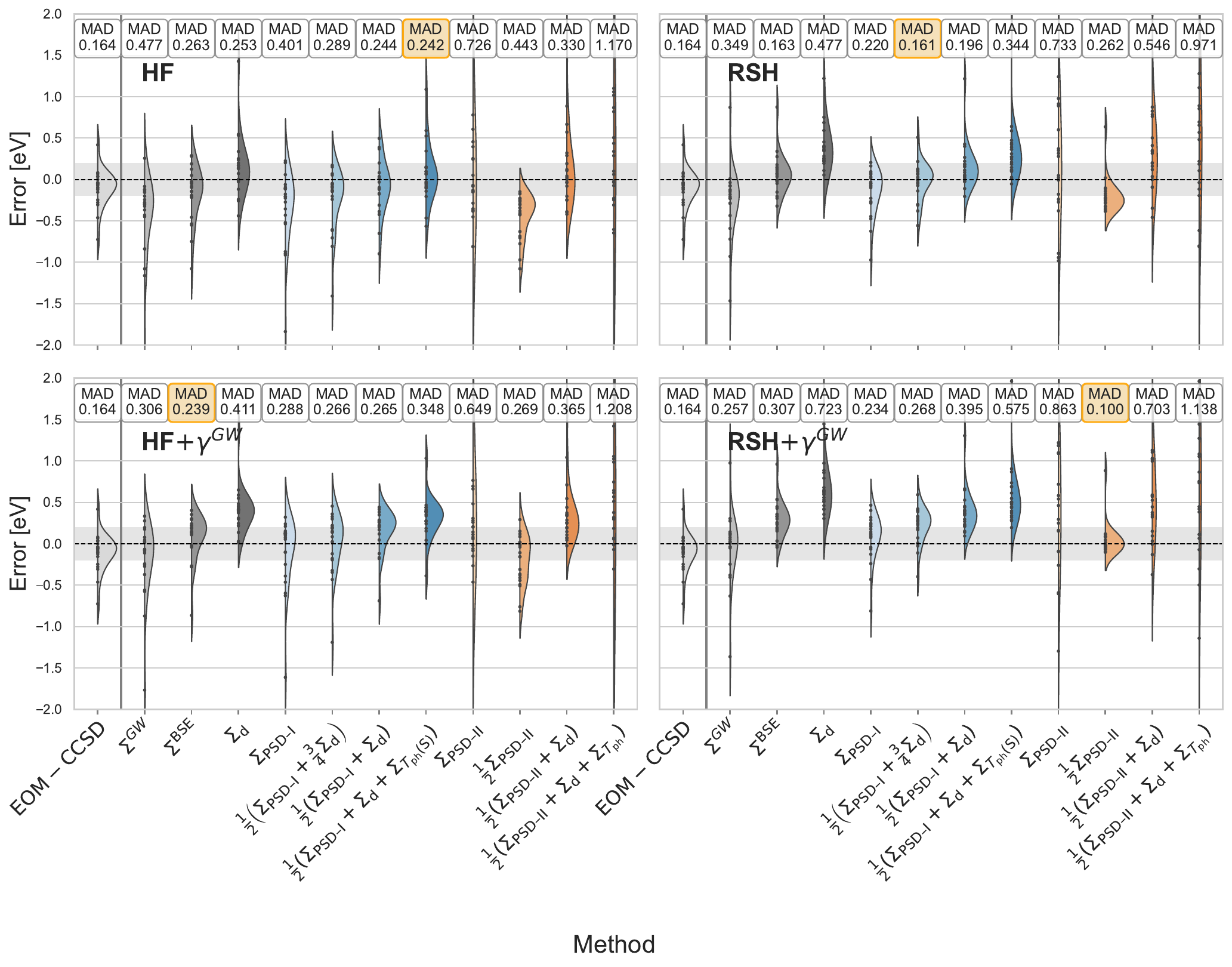}
  \begin{caption}{\small Error distribution of the HOMO-1 energies with respect to a reference CCSD(T) $\Delta$SCF evaluation for the 23 molecules contained in the Marie and Loos set~\cite{marie_reference_2024}. Different self-energies are evaluated as one-shot perturbations on top of four different reference states.
      \label{fig:bench:homo-1}}
\end{caption}
\end{figure}
The new self-energies we have introduced in this section have appealing formal properties. However, the true test is the actual accuracy of these approximations in real molecular systems. As the first benchmark we use the set of molecules by Marie and Loos~\cite{marie_reference_2024}.

This benchmark set comprises 23 small molecules systematically grouped by valence and isoelectronic character: the 10-electron (Ne, HF, H$_2$O, NH$_3$, CH$_4$), 12-electron (LiF, BeO, BN, C$_2$), 14-electron (N$_2$, CO, BF), and 18-electron (CS, Ar, HCl, H$_2$S, PH$_3$, SiH$_4$, LiCl) series, alongside a subset of miscellaneous molecules (F$_2$, CO$_2$, CH$_2$O, BH$_3$). All calculations in the original benchmark employ geometries optimized at the CC3/aug-cc-pVTZ level of theory without the frozen-core approximation.  The reference energies provided in this dataset are of near full configuration interaction (FCI) quality, having been computed using the Configuration Interaction using a Perturbative Selection made Iteratively (CIPSI) method.\cite{Huron1973} In our calculations, we omit BN and C$_2$ since they do not have a stable singlet ground state at the mean-field level. 

Our results for quasiparticle HOMO (Fig.~\ref{fig:bench:homo}) and HOMO-1 (Fig.~\ref{fig:bench:homo-1}) energies demonstrate a systematic error sensitivity to the initial reference state: Hartree-Fock (HF) vs. optimally tuned range-separated hybrid exchange-correlation functional~\cite{refaely-abramson_quasiparticle_2012} (RSH) using the scheme proposed in Ref.~\citenum{mckeon_optimally_2022}. Additionally, we tested the impact of evaluating the static self-energy contribution with an improved correlated density matrix at the linearized $GW$ level (denoted as $\gamma^{\text{GW}}$).\cite{Bruneval2019, Bruneval2019a} This yields four reference states in total (HF, RSH, HF$+\gamma^{\text{GW}}$, and RSH$+\gamma^{\text{GW}}$), on top of which all self-energies are evaluated as one-shot perturbative corrections. Errors are measured with respect to the CIPSI reference ionization energies; the EOM-CCSD column, evaluated against the same reference, indicates the accuracy attainable by relatively affordable single-reference coupled-cluster theory and thus sets a natural target for the methods under study.

\begin{figure}[t!]
  \center
  \includegraphics[trim=0cm 0.75cm 0cm 0cm, clip, width=0.82\textwidth]{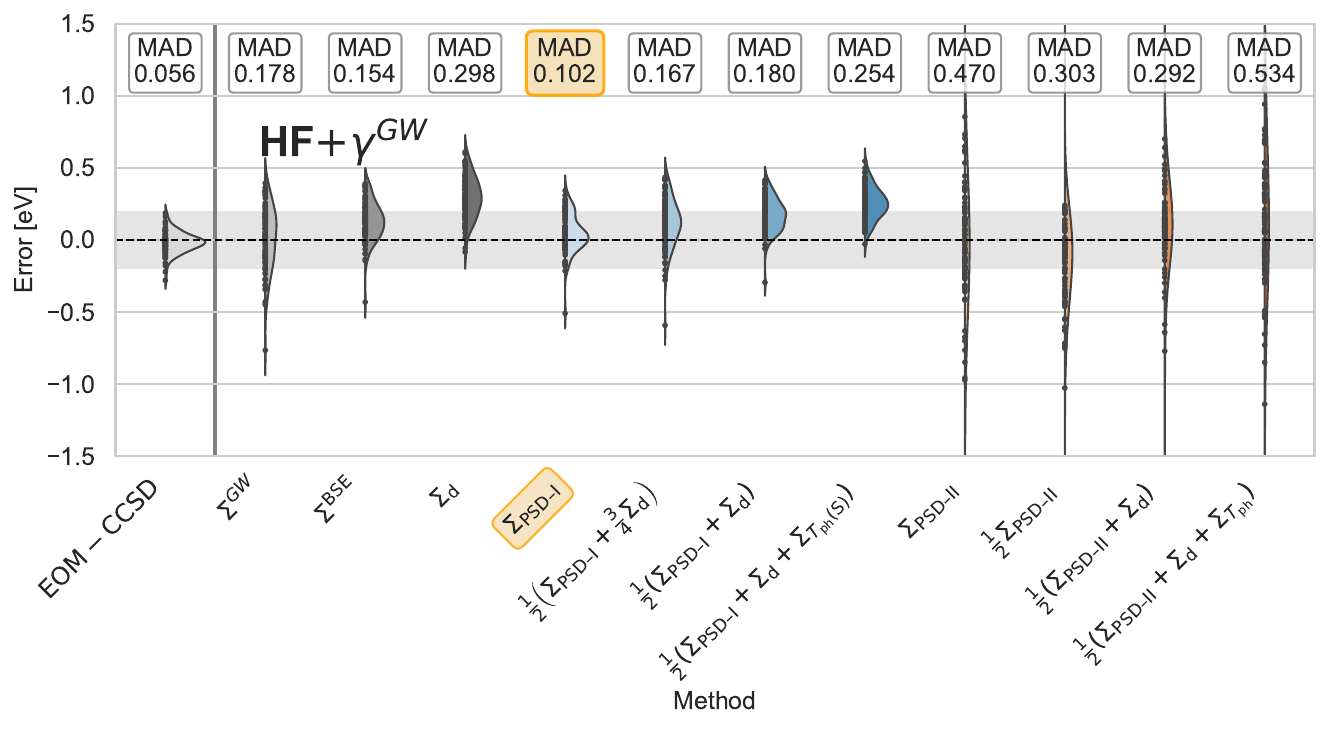}
  \begin{caption}{\small Error distribution of the HOMO energies with respect to a reference CCSD(T) $\Delta$SCF evaluation for the 100 molecules contained in the GW100 set~\cite{van_setten_gw_2015}.
      \label{fig:bench:gw100}}
\end{caption}
\end{figure}

Several clear trends emerge from Fig.~\ref{fig:bench:homo}. On top of the HF reference, the $GW$ approximation performs worst, with a mean absolute deviation (MAD) of 0.43~eV. Upgrading the screening from the RPA to the BSE level reduces this error by more than a third, whether through the original self-energy of Ref.~\citenum{forster_why_2024} ($\Sigma^{\text{BSE}}$) or already through the direct term $\S_d$ alone; the best synthetic combination, $\frac12(\S_{\text{PSD-I}}+\S_d+\S_{T_{ph}(S)})$, brings it down to 0.25~eV. The improvement becomes considerably more pronounced for better reference states. With the RSH starting point, $\S_{\text{PSD-I}}$ (MAD of 0.14~eV) clearly outperforms $GW$ (0.24~eV), and the admixture $\frac12(\S_{\text{PSD-I}}+\frac34\S_d)$ reaches 0.13~eV. The best overall performance is achieved when the static contribution to the self-energy is evaluated using the $GW$ linearized density matrix. For HF$+\gamma^{\text{GW}}$, the minimal PSD self-energy attains a MAD of 0.095~eV, remarkably close to the intrinsic accuracy of EOM-CCSD (0.082~eV). This is in agreement with previous studies that demonstrated substantial improvements of ionization potentials at the $GW$ level\cite{bruneval_gw_2021} as well as with vertex-corrected approaches.\cite{Tolle2026ConnectionClusterb}

Notably, $\gamma^{\text{GW}}$ also attenuates the starting-point dependence of ${\text{PSD-I}}$: while its MAD differs by a factor of two between the HF and RSH references, the two density-matrix-corrected references yield 0.095 and 0.13~eV, respectively.

The PSD-II family, which admits intermediate triplet excitations, shows the opposite behavior: for all four references it degrades the quasiparticle energies, with MADs typically between 0.2 and 0.5~eV and broad error distributions extending far beyond those of the singlet-only schemes. This behavior can be traced back to the known difficulty of the statically screened BSE kernel in the triplet channel, where excitation energies are systematically underestimated~\cite{jacquemin_is_2017, Rangel2017, blase_bethesalpeter_2018}. Since the triplet states enter $\S_{\text{PSD-II}}$ with a prefactor of three [Eq.~\eqref{eq:PSD-II}], even moderate errors in the triplet manifold translate into a sizable spurious correlation contribution. The PSD construction itself is agnostic to the spin multiplicity of the intermediate states; the benchmark thus indicates that restricting the sum to singlets, i.e., the minimal completion $\S_{\text{PSD-I}}$, is not only the simplest but also the most accurate strategy as long as the underlying triplet excitation energies are not improved.

The HOMO-1 energies shown in Fig.~\ref{fig:bench:homo-1} probe the theory one step deeper in the valence region, where stronger correlation effects and incipient satellite features come into play,\cite{Cederbaum1977} and the performance of most methods that perform well for the first IP deteriorates.\cite{marie_reference_2024} The errors increase for all methods; the MAD of EOM-CCSD itself doubles to 0.16~eV, so part of the deterioration reflects the increased difficulty of these states rather than a failure of any particular self-energy. The trends established for the HOMO largely persist: the PSD-I-based schemes outperform $GW$ for every reference, with $\frac12(\S_{\text{PSD-I}}+\frac34\S_d)$ on the RSH reference (0.16~eV) matching the accuracy of EOM-CCSD, although their advantage over the parent $\Sigma^{\text{BSE}}$ self-energy is less systematic than for the HOMO. We also note the conspicuously low MAD of $\frac12\S_{\text{PSD-II}}$ on the RSH$+\gamma^{\text{GW}}$ reference (0.10~eV); given the strong reference sensitivity of the PSD-II family elsewhere, we attribute it to a fortuitous error cancellation rather than to a systematic improvement.

The Marie--Loos set is invaluable for its FCI-quality reference energies, but it comprises only 23 light molecules. To verify that our conclusions are not an artifact of this selection, we took the best-performing combination of mean field reference and static self-energy from the first benchmark ($\text{HF}+\gamma^{\text{GW}}$) and repeated the comparison on the larger GW100 set of 100 small- and medium-sized molecules~\cite{van_setten_gw_2015} using the improved reference values at the $\Delta$CCSD(T) level of theory of Ref.~\citenum{bruneval_gw_2021}. These results are presented in Fig.~\ref{fig:bench:gw100}. The picture established above carries over with remarkable consistency: the MAD of $\S_{\text{PSD-I}}$ (0.10~eV) is essentially unchanged with respect to the smaller set, reducing the $GW$ error (0.18~eV) by almost a factor of two and improving markedly on the parent $\Sigma^{\text{BSE}}$ self-energy (0.15~eV). It also outperforms $GW$@RSH, which has so far achieved the lowest MAD out of all $GW$-based methods tested on GW100.\cite{mckeon_optimally_2022} Beyond the MAD, the error distribution of $\S_{\text{PSD-I}}$ is narrow, nearly symmetric, and centered at zero, whereas $\S_d$ and the synthetic mixtures acquire a systematic positive bias, and the PSD-II variants again develop long error tails. The relative ordering of the approximations is preserved between the two benchmark sets, lending confidence to the transferability of the singlet-only PSD correction. Taken together, the two benchmarks single out $\S_{\text{PSD-I}}$ evaluated on a density-matrix-corrected reference as the recommended scheme: it restores the positivity of the spectral function and delivers quasiparticle energies within 0.1~eV of coupled-cluster quality at essentially the same computational cost as the original BSE-based self-energy.

\section*{Discussion}
We have derived positive semidefinite extensions of the BSE-based electron self-energy recently introduced in Ref.~\citenum{forster_why_2024}. The parent scheme is accurate and numerically efficient; like virtually all diagrammatic approximations incorporating vertex corrections consistently in the 2-particle correlation function $L$ and self-energy,\cite{Kutepov2016, Kutepov2017a, maggio_gw_2017, Vlcek2019, Vacondio2024, Patterson2024MolecularCorrections, Forster2025} however, it admits small regions of negative spectral weight~\cite{stefanucci_diagrammatic_2014}. Working within the Non-equilibrium Green's function (NEGF) framework of Refs.~\citenum{stefanucci_diagrammatic_2014, uimonen_diagrammatic_2015}, we partitioned its exchange-type diagram into half-diagrams on the Keldysh contour and completed them to a full square. This construction is not a mere regularization: the diagrams generated along the way describe genuine scattering processes of the particle-hole $T$-matrix type, and thereby mix correlation channels that are usually combined \emph{ad hoc}. The resulting family of PSD self-energies is parametrized by the spin multiplicity of the intermediate BSE excitations and by admixtures restoring the exact $GW$, PT2, and $T_{ph}$ limits.

Extensive benchmarks on the Marie--Loos\cite{marie_reference_2024} and GW100\cite{van_setten_gw_2015} sets identify the minimal, singlet-only completion $\S_{\text{PSD-I}}$, evaluated on a density-matrix-corrected Hartree-Fock reference, as the method of choice. It guarantees a nonnegative rate function, converts the small amounts of negative spectral weight into physical satellite channels, and reduces the mean absolute deviation of HOMO energies to about 0.1~eV, close to the intrinsic accuracy of EOM-CCSD, while retaining the relative computational efficiency of the parent scheme. Admitting triplet intermediate states, by contrast, degrades the results, reflecting the well-documented weakness of the statically screened BSE kernel in the triplet channel~\cite{jacquemin_is_2017}; any progress on that front would immediately carry over to the corresponding PSD self-energies. Natural extensions of this work include self-consistent evaluations, the analysis of satellite spectra in comparison with cumulant techniques~\cite{pavlyukh_vertex_2016}, and applications to extended systems, where the benefits of excitonic screening are well-established\cite{Kutepov2016, Cunningham2018, Cunningham2023, Koskelo2025Short-RangeMetals} but its interplay with the PSD construction remains unexplored.
\section*{Methods}
\label{sec:methods}
\subsection*{Functional-derivative derivation of the self-energy}
\label{sec:derivation}
Equation~\eqref{eq:se} follows from Schwinger's functional-derivative technique~\cite{hedin_new_1965, strinati_application_1988, stefanucci_nonequilibrium_2025}; integration over repeated arguments is implied. In the presence of an auxiliary local external potential $U(\var{3})$, set to zero at the end of the derivation, the equation of motion for the GF casts the exact self-energy (Hartree plus exchange-correlation, Hxc) in the Schwinger-Dyson form
\begin{align}
  \S_{\text{Hxc}}(\var{1}, \var{2}) &= i v(\var{1}^+, \var{3})\, G(\var{1}, \var{4})\, \tilde{\Gamma}(\var{4},\var{2};\var{3}), &
  \tilde{\Gamma}(\var{4},\var{2};\var{3}) &= -\frac{\delta G^{-1}(\var{4},\var{2})}{\delta U(\var{3})},
  \label{eq:sde}
\end{align}
where $\tilde{\Gamma}$ is the three-point vertex function. Substituting the Dyson equation $G^{-1} = G_0^{-1} - U - \S_{\text{Hxc}}$ splits the vertex into a trivial part and a correlation part:
\begin{align}
  \tilde{\Gamma}(\var{4},\var{2};\var{3}) &= \delta(\var{4},\var{2})\delta(\var{4},\var{3})
  + \frac{\delta \S_{\text{Hxc}}(\var{4},\var{2})}{\delta U(\var{3})}.
  \label{eq:vertex}
\end{align}
In Eq.~\eqref{eq:sde}, the trivial part generates the Fock exchange $i v(\var{1}^+, \var{2}) G(\var{1}, \var{2})$, which belongs to the mean-field reference. The remaining term carries all correlations, and the chain rule resolves it into the two building blocks of our theory:
\begin{align}
  \frac{\delta \S_{\text{Hxc}}(\var{4},\var{2})}{\delta U(\var{3})}
  = \frac{\delta \S_{\text{Hxc}}(\var{4},\var{2})}{\delta G(\var{5},\var{6})}\,
  \frac{\delta G(\var{5},\var{6})}{\delta U(\var{3})}
  = I(\var{4},\var{6},\var{2},\var{5})\, L(\var{5},\var{3},\var{6},\var{3}),
  \label{eq:chain}
\end{align}
where, in accordance with the definition below Eq.~\eqref{eq:se}, we identified the particle-hole irreducible four-point kernel and the two-particle correlation function with degenerate arguments:
\begin{align}
  \frac{\delta \S_{\text{Hxc}}(\var{4},\var{2})}{\delta G(\var{5},\var{6})} &= -i I(\var{4},\var{6},\var{2},\var{5}), &
  \frac{\delta G(\var{5},\var{6})}{\delta U(\var{3})} &= i L(\var{5},\var{3},\var{6},\var{3}).
  \label{eq:ident}
\end{align}
Substituting Eq.~\eqref{eq:chain} into Eq.~\eqref{eq:sde} yields Eq.~\eqref{eq:se}. The same ingredients determine $L$ itself: applying $\delta/\delta U$ to the Dyson equation gives $\delta G/\delta U = G\,\tilde{\Gamma}\,G$, which together with Eqs.~\eqref{eq:vertex} and \eqref{eq:chain} reproduces the BSE~\eqref{eq:bse}. Consistency therefore demands that the same kernel enters Eq.~\eqref{eq:se} and the BSE. Finally, evaluating $I$ for the self-energy $\S = \S_H + iGW_0$ with a fixed statically screened interaction recovers Eq.~\eqref{eq:kr:L}: the Hartree term contributes the bare interaction $v$, and the screened exchange contributes $-W_0$.
\subsection*{Spin-resolved Casida equation}
\label{sec:casida}
The excitation energies $\Omega_\nu$ and transition amplitudes $\vec{Z}^\nu$ entering the half-diagrams are obtained from the Casida secular problem~\cite{casida_time-dependent_2009} in the spin-orbital basis,
\begin{align}
\begin{pmatrix} \mat{A} & \mat{B} \\ -\mat{B} & -\mat{A} \end{pmatrix}
\begin{pmatrix} \vec{X}^\nu \\ \vec{Y}^\nu \end{pmatrix}
= \Omega_\nu \begin{pmatrix} \vec{X}^\nu \\ \vec{Y}^\nu \end{pmatrix},
\label{eq:casida}
\end{align}
with the excitation (deexcitation) amplitudes $\vec{X}^\nu_{i\sigma_i,a\sigma_a}$ ($\vec{Y}^\nu_{i\sigma_i,a\sigma_a}$) normalized as $\sum_{ia}\sum_{\sigma_i\sigma_a}(|\vec{X}^\nu_{i\sigma_i,a\sigma_a}|^2 - |\vec{Y}^\nu_{i\sigma_i,a\sigma_a}|^2)=1$. For the kernel~\eqref{eq:kr:L} and a restricted closed-shell reference with real orbitals, the coupling matrices are conveniently expressed through
\begin{align}
  \mat{\Delta}_{ia,jb} &= (\epsilon_a{-}\epsilon_i)\delta_{ij}\delta_{ab}, &
  \mat{J}_{ia,jb} &= (ia|v|jb), &
  \mat{K}_{ia,jb} &= (ij|W_0|ab), &
  \mat{K}'_{ia,jb} &= (ib|W_0|aj),
  \label{eq:JK}
\end{align}
where $\epsilon_p$ are the quasiparticle energies of the underlying reference. The direct integrals involve the bare interaction, while the exchange-type integrals involve the statically screened one.

Spin symmetry reduces the problem substantially. Because the kernel conserves the spin projection, the spin-conserving ($M_s=0$) and spin-flip ($M_s=\pm1$) sectors decouple. In the $M_s=0$ sector, the unitary combinations $\vec{X}^{S/T_0}=\tfrac{1}{\sqrt{2}}(\vec{X}_{\a\a}\pm\vec{X}_{\b\b})$ (and analogously for $\vec{Y}$) block-diagonalize Eq.~\eqref{eq:casida} into a singlet and a triplet problem of the same form with
\begin{subequations}
  \label{eq:casida:ST}
\begin{align}
  \mat{A}_S &= \mat{\Delta} + 2\mat{J} - \mat{K}, & \mat{B}_S &= 2\mat{J} - \mat{K}',\\
  \mat{A}_T &= \mat{\Delta} - \mat{K}, & \mat{B}_T &= -\mat{K}'.
\end{align}
\end{subequations}
The two spin-flip blocks are governed by the same matrices $\mat{A}_T$ and $\mat{B}_T$; in the absence of spin-orbit coupling, the three triplet components are therefore exactly degenerate and share identical orbital amplitudes, $\vec{X}^{T_0}=\vec{X}^{T_{\pm1}}$ and $\vec{Y}^{T_0}=\vec{Y}^{T_{\pm1}}$. Collecting the amplitudes into the matrix $\vec{Z}^\nu_{ia}=\vec{X}^\nu_{ia}$, $\vec{Z}^\nu_{ai}=\vec{Y}^\nu_{ia}$, these observations are summarized by the spin symmetries
\begin{align}
   \begin{cases}
  \vec{Z}^\nu_{p\a,q\a} = +\vec{Z}^\nu_{p\b,q\b}\equiv \tfrac{1}{\sqrt{2}}\vec{Z}^\nu_{pq}, & S_\nu=0,\; S^z_\nu=0,\\
  \vec{Z}^\nu_{p\a,q\a} = -\vec{Z}^\nu_{p\b,q\b}\equiv \tfrac{1}{\sqrt{2}}\vec{Z}^\nu_{pq}, & S_\nu=1,\; S^z_\nu=0,\\
  \vec{Z}^\nu_{p\a,q\b} = \vec{Z}^\nu_{p\b,q\a} \equiv \vec{Z}^\nu_{pq}, & S_\nu=1,\; S^z_\nu=\pm1,
  \end{cases}\label{eq:Z:spin}
\end{align}
which lead directly to the simplified half-diagrams~\eqref{half:diag:expl}. The complete block-by-block derivation, the verification of the spin purity of the solutions through the expectation value of $\hat{S}^2$, and a comparison with the unified spin-adaptation scheme of Ref.~\citenum{angyan_correlation_2011} are provided in the Supplementary Information.

The spin structure also explains why only singlet states enter the self-energies $\S_d$ and $\S_{2x}$ [Eqs.~\eqref{eq:Sigma:d} and \eqref{eq:Sigma:x}]. The spin-summed transition density matrix of the excited state $\nu$ projects exclusively onto its singlet components: the density operator is spin-diagonal, so the spin-flip amplitudes drop out, while the explicit spin summation cancels the antisymmetric $M_s=0$ triplet combination,
\begin{align}
  \delta\rho_\nu(r_1, r_2) &= \sum_{\tau\in\{\a,\b\}}\langle0|\hat{\psi}^\dagger_\tau(r_2)\hat{\psi}_\tau(r_1)|\nu\rangle
  = \sqrt{2}\sum_{ia}\left[\vec{X}^{S,\nu}_{ia}\varphi_a(r_1)\varphi^*_i(r_2) + \vec{Y}^{S,\nu}_{ia}\varphi_i(r_1)\varphi^*_a(r_2)\right],
  \label{eq:trans:dens}
\end{align}
where $\hat{\psi}_\tau(r)=\sum_p \varphi_p(r)\,\hat{a}_{p\tau}$ is the field operator annihilating an electron with spin $\tau$ at point $r$, and $\hat{a}_{p\tau}$ is the fermionic annihilation operator for orbital $\varphi_p$. Since triplet eigenvectors have vanishing singlet components, $\delta\rho_\nu$ vanishes identically for all triplet excitations, and with it the half-diagram $a^\nu$~\eqref{a:diag:expl}.

\subsection*{Diagrammatic method for half-diagrams}
\label{sec:diagrams}
We now summarize how the half-diagrams and the rules for their recombination follow from the general PSD theory of Ref.~\citenum{stefanucci_diagrammatic_2014}. Cutting the diagram in Fig.~\ref{fig:3} along the border between the islands of pluses and minuses severs the intermediate lesser Green's function and the excited-state propagator; the two resulting halves read, in terms of contour quantities,
\begin{subequations}
  \label{eq:half:diag:contour}
\begin{align}
  a^\nu_{\scp_4}(\var{1}) &= \int d\scr_3\, v(\scr_1, \scr_3)\, g^<(\var{1},\scp_4)\, \chi^{<,\nu}(\scr_3,\scr_3),\\
  b^\nu_{\scp_4}(\var{2}) &= -\int d\scr_4\, W_0(\scr_2, \scr_4)\, g^<(\scp_4,\var{4})\, \chi^{<,\nu}(\scr_2,\scr_4),
\end{align}
\end{subequations}
where $\scr_i=(r_i,s_i)$ comprises position and spin, $\var{4}=(\scr_4, t_2)$ because $W_0$ is instantaneous, and the excited-state-resolved response function
\begin{align}
  \chi^{<,\nu}(\scr,\scr')&= \sum_{\scp\scq} \vec{Z}^\nu_{\scp\scq}\,\varphi_{\scp}^*(\scr)\varphi_{\scq}(\scr')
  \label{eq:chi:nu}
\end{align}
follows from the Lehmann representation of $L^<$. Both halves depend on their time arguments only through phase factors set by the intermediate energies $\epsilon_k$ and $\Omega_\nu$; stripping these phases and projecting on molecular orbitals yields the time-independent amplitudes
\begin{subequations}
  \label{eq:ab:general}
\begin{align}
  a^\nu_{\scp_4}(\scp_1) & = \sum_{\scp_3\scq_3} v_{\scp_1\scq_3\scp_3\scp_4}\, \vec{Z}^\nu_{\scp_3\scq_3},\\
  b^\nu_{\scp_4}(\scq_2) & = -\sum_{\scp_2\scq_4} W^0_{\scp_2\scp_4\scq_4\scq_2}\, \vec{Z}^\nu_{\scp_2\scq_4},
\end{align}
\end{subequations}
which, after performing the spin sums, reduce to Eqs.~\eqref{eq:ab:matel}. Gluing two halves back together [Eqs.~(21), (22), and (24) of Ref.~\citenum{stefanucci_diagrammatic_2014}] restores the lesser self-energy as a sum of products of two amplitudes carrying the common phase $e^{-i(\epsilon_k-\Omega_\nu)(t_1-t_2)}$: each pair of half-diagrams describes the rate of a scattering process in which a hole at energy $\epsilon_k$ has shaken up the excitation $\Omega_\nu$. Fourier transformation and the reconstruction of the time-ordered component convert these rates into the resolvents $(\w-\epsilon_k+\Omega_\nu-i\h)^{-1}$ of Eqs.~\eqref{eq:Arno} and \eqref{eq:PSD-II:expl}; the greater component is treated identically, with unoccupied intermediate states and poles at $\w=\epsilon_c+\Omega_\nu$.

In this language, the diagrams of the original theory correspond to the ordered pairs of half-diagrams $(a,a)\rightarrow\S_d$ and $(a,b)\rightarrow\S_{2x}$ (Fig.~\ref{fig:1}). A self-energy is PSD when its constituent pairs form a complete square $\tilde{I}_1\times\tilde{I}_1$ over some set $\tilde{I}_1$ of half-diagrams [Eq.~(33) of Ref.~\citenum{stefanucci_diagrammatic_2014}]. The pair set $\{(a,a), (a,b)\}$ is not of this form; its minimal completion is $\tilde{I}_1=\{a, b\}$, whence
\begin{align}
  \left(a+b,\, a+b\right) &= (a,a) + (a,b) + (b,a) + (b,b),
  \label{eq:complete:square}
\end{align}
which is precisely the structure of Eq.~\eqref{eq:psd:gen}: the exchange term is doubled, and the particle-hole $T$-matrix term $(b,b)$ emerges. Because $W_0$ is distinct from $v$, all pairs are topologically distinct and no double counting arises at any order. For $W_0=v$, in contrast, the lowest-order $a$ and $b$ halves coincide up to a permutation of the dangling lines, the completion must additionally include permutations of the intermediate lesser lines, and the resulting square generates further classes of diagrams in which the two-particle correlation function enters with non-degenerate time arguments (three independent frequencies). This is the technical reason for excluding the TDHF scenario in the ``Ingredients'' subsection.

\subsection*{Computational Details}
We have implemented all the methods in this work into the BAND engine of a modified development version of the Amsterdam modelling suite (AMS)\cite{Baerends2025} as well as MOLGW.\cite{Bruneval2016a} All results in this work have been produced using MOLGW, in the aug-cc-pvQZ basis set, the same basis set at which the reference results have been calculated. For the optimally-tuned RSH calculations, we adopted the prescription of Ref.~\citenum{mckeon_optimally_2022} that we implemented through the LibXC library.\cite{Lehtola2018} The optimal tuning parameters have been determined for each self-energy approximation separately.

\section*{Data availability}
The source data supporting the findings of this study are available via this article and its supplemental material.

\section*{Code availability}
The code used to perform all calculations is available from the authors upon reasonable request, and will be released in the public version of MOLGW under 
\href{https://github.com/molgw/molgw}{https://github.com/molgw/molgw}

\section*{Acknowledgements}
AF acknowledges funding through a VENI grant from NWO under grant agreement VI.Veni.232.013.
This work was performed using HPC resources from GENCI-CCRT-TGCC (Grants No. 2025-096018).

\section*{Supplementary Information}
The following file is available free of charge.
\begin{itemize}
  \item Supplementary notes (PDF): complete block-by-block derivation of the spin-resolved Casida equation; verification of the spin purity and degeneracy of the excitation manifold through the expectation value of $\hat{S}^2$, including a comparison with the unified spin-adaptation scheme of Ref.~\citenum{angyan_correlation_2011}; demonstration that only singlet excitations contribute to the self-energies $\S_d$ and $\S_{2x}$.
\end{itemize}

\section*{Author contributions}
All authors conceptualized the work together. YP took the lead in formulating the theoretical framework and wrote the initial draft. FB and AF independently implemented the approach and validated their implementations. FB performed all calculations. All authors contributed to the final reviewing and editing.
\section*{Competing interests}
There are no competing interests to declare.

\bibliography{MyLibrary.bib,references.bib}

\newpage

\begin{center}
\begin{minipage}[b][1.75in]{3.25in}
  \centering
  \includegraphics[width=3.25in,height=1.75in,keepaspectratio]{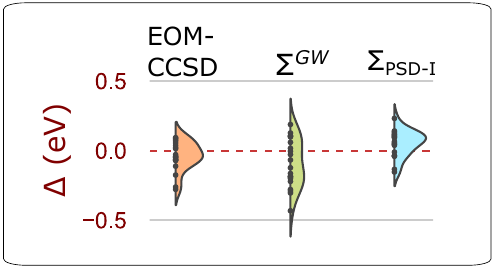}
\end{minipage}
\end{center}

\end{document}